\documentclass{ws-procs11x85}

\usepackage{ws-procs-thm}           
\usepackage{booktabs}
\usepackage{siunitx} 

\newcommand{\R}{\mathbb{R}}

\DeclareMathOperator*{\argmax}{arg\,max}

\begin{document}

\title{Cross-modal representation alignment of molecular structure and perturbation-induced transcriptional profiles}

\author{Samuel G. Finlayson$^{1,2,*}$, Matthew B.A. McDermott$^{2,*}$, Alex V. Pickering$^3$, Scott L. Lipnick$^3$, Isaac S. Kohane$^{3,\dag}$ }


\address{$^1$Department of Systems, Synthetic, and Quantitative Biology, Harvard Medical School, Boston, MA \\
$^2$Department of EECS, Massachusetts Institute of Technology, Cambridge, MA
\\
$^3$Department of Biomedical Informatics, Harvard Medical School, Boston, MA
\\
$^*$Co-first author
\\
$^\dag$E-mail: isaac\_kohane@harvard.edu}

\begin{abstract}
Modeling the relationship between chemical structure and molecular activity is a key goal in drug development. Many benchmark tasks have been proposed for molecular property prediction, but these tasks are generally aimed at specific, isolated biomedical properties. In this work, we propose a new cross-modal small molecule retrieval task, designed to force a model to learn to associate the structure of a small molecule with the transcriptional change it induces. We develop this task formally as multi-view alignment problem, and present a coordinated deep learning approach that jointly optimizes representations of both chemical structure and perturbational gene expression profiles. We benchmark our results against oracle models and principled baselines, and find that cell line variability markedly influences performance in this domain. Our work establishes the feasibility of this new task, elucidates the limitations of current data and systems, and may serve to catalyze future research in small molecule representation learning.
\end{abstract}

\copyrightinfo{\copyright\ 2020 The Authors. Open Access chapter published by World Scientific Publishing Company and distributed under the terms of the Creative Commons Attribution Non-Commercial (CC BY-NC) 4.0 License.}

\section{Introduction}
Identifying molecules that are likely to have a specific biological effect is a cornerstone of drug discovery and a key component of efforts to achieve precision medicine. Classically, computational profiling of small molecules has centered on predicting affinities for specific biological targets, using tools ranging from biophysics-driven techniques such as molecular docking\cite{hansson2002molecular} to literature-mined annotations.\cite{krallinger2017information} Small molecule modeling has also recently become a major area of interest in deep learning, a trend catalyzed by graph neural networks \cite{battaglia2018relational} and benchmarking datasets.\cite{wu2018moleculenet} Graph neural networks allow for end-to-end modeling of molecular graphs,\cite{kearnes2016molecular, yang2019analyzing, lo2018machine, ramsundar2015massively} and have yielded state-of-the-art performance on certain tasks.\cite{faber2017prediction, gilmer2017neural} In addition, deep learning approaches have been used at a more global scale modeling cross-molecule relationships.\cite{zitnik2018modeling}
To date, deep learning efforts in this space have generally focused on two extremes: highly local, biochemical prediction problems, which test the model's ability to predict specific chemical properties, and more global, clinical modeling tasks, such as indication or side effect prediction. 
Missing from the field, however, are benchmark tasks between these two extremes, that test the ability of deep models to encode rich, general representations of a molecule's broad-spectrum effect on cellular biology.

In parallel to these developments, connectivity mapping has emerged as a alternative approach for drug development\cite{musa2017review}. In connectivity mapping, compounds are foremost characterized not by individual chemical properties or downstream targets, but by the broad transcriptional effects they induce in cells. Connectivity mapping begins by first developing a large dataset of \emph{perturbational signatures} of  molecules by physically treating cell lines with these molecules, then measuring the resultant changes in gene expression. These datasets are then compared to one or more \emph{query signatures}, which are typically differential gene expression (GE) signatures representing disease states that investigators hope to reverse. Various public datasets have been curated to enable these efforts,\cite{lamb2006connectivity,subramanian2017next} and researchers have sought to use these for drug repurposing, precision medicine, and analysis of gene expression data in general.\cite{musa2018harnessing,liu2018systematic,clark2014characteristic,donner2018drug,dincer2018deepprofile,rampavsek2019dr,cheng2013evaluation,lopez2018deep}

Connectivity mapping is promising because it can be used to search for new indications of drugs without making any specific \emph{a priori} assumptions about their mechanism of action.  However, the typical framework for connectivity mapping is limited by the fact that it can only query against drugs that have already been profiled using the transcriptional assay. In other words, connectivity mapping is -- in principle -- very flexible with respect to the disease signatures they accept as a query, but is transductive rather than inductive with respect to the target small molecule signatures.
This is the perfect complement to structure-based computational chemistry, which is typically inductive to new drug structures but can only make predictions for diseases with known targets.

\begin{figure*}
\begin{center}
\includegraphics[width=0.85\linewidth]{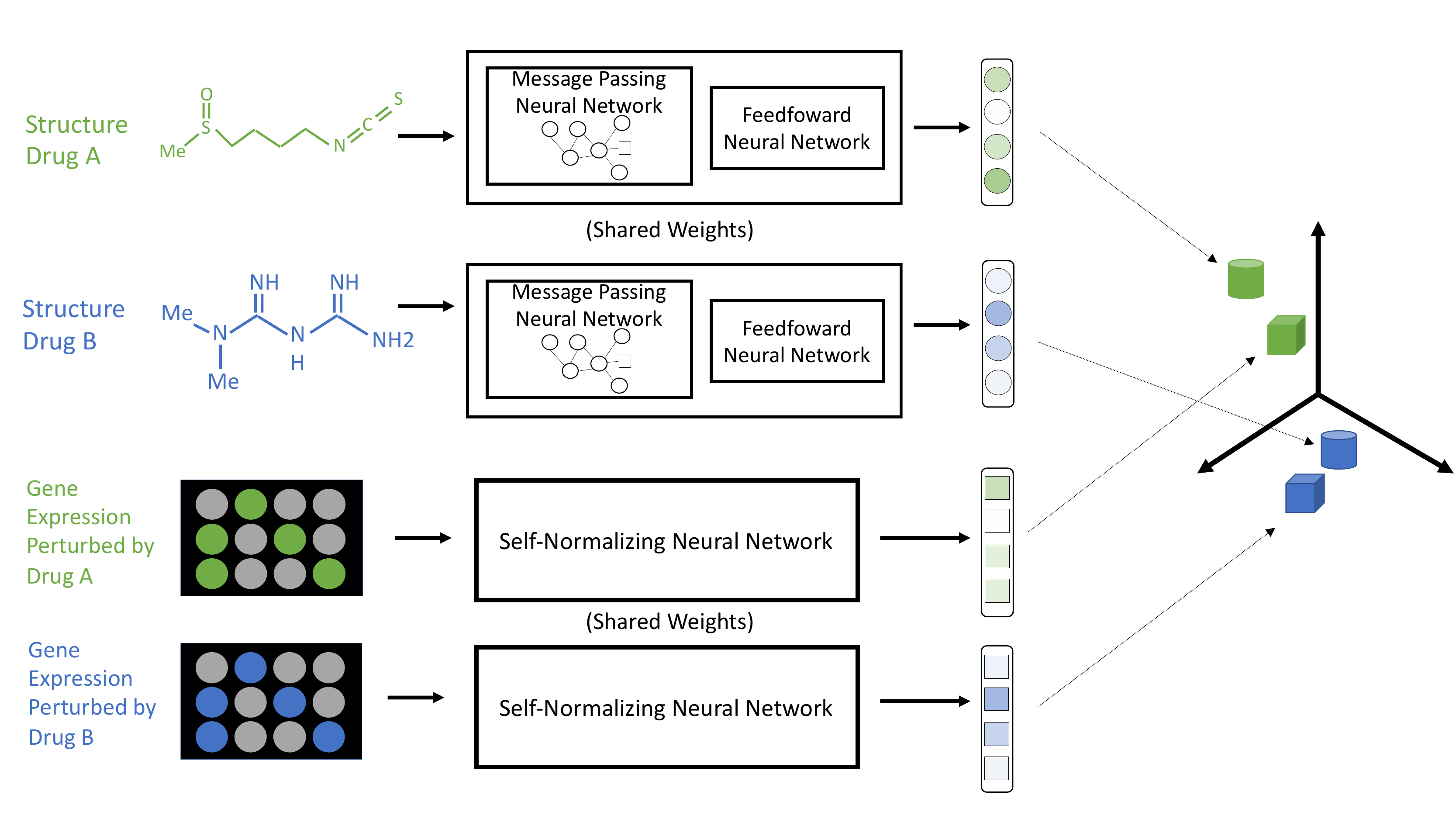}
\end{center}
\caption[Aligned Chemical-Transcriptomic Representation Learning Method]{Our representation learning method. Neural networks are trained to embed gene expression profiles close to the small molecule structures that induce them. Given a cross-modal alignment, gene expression signatures can be used as queries to rank chemical structures by their likelihood to induce such a signature.}
\label{aba:fig1}
\end{figure*}

In this work, we combine these two fields, by using deep chemical embedders to learn the transcriptional space encoded by CMAP profiling. More specifically, we train coordinated networks to jointly embed chemical structures and perturbational gene expression profiles such that learned chemical representations are most similar to the encodings of the transcriptional patterns they induce.\footnote{Code available here: \url{https://github.com/sgfin/molecule_ge_coordinated_embeddings}}. Note that this task naturally fills the gap in inductive molecular modelling identified previously; by tasking the model to produce highly similar embeddings for chemical structures and the perturbational profiles they induce, we force the model to learn a  \emph{transcriptome-wide} reflection of the drug's action on the cell. We then evaluate these chemical representations by using gene expression signatures as queries into the embedding space and recovering their corresponding compounds. (See Figure~\ref{aba:fig1}). Crucially, the evaluation is set up such that the validation and test set compounds and cell lines are not used in training, which allows us to test the ability of the model to generalize to new drugs and cell lines. 


In the rest of this work, we first offer some background on joint embedding alignment, then detail the methods used in this work.
Finally, we walk through the results and discussion of these experiments, then close with concluding thoughts. A version of this work, with supplementary material present, can be found here: \url{https://github.com/sgfin/molecule_ge_coordinated_embeddings/blob/master/paper_08_2020.pdf}.

\section{Background}
In \emph{multi-view representation alignment}, embeddings of two associated data modalities are learned separately but in a coordinated manner, such that the resulting embeddings are similar.
These methods have been used in comparing images to text but also in other domains.\cite{hsu2018unsupervised,hassani2020contrastive}
In this work, we learn aligned representations such that small molecules are embedded in close proximity to the differential gene expressions they induce. Multi-view representation alignment can be achieved through a variety of methods, including classical methods, such as \emph{canonical correlation analysis} (CCA)\cite{hotelling1992relations} and methods using distance, similarity, correlation, or ranking based penalties during training.\cite{li2018survey} 
Ranking-based methods for multi-view representation alignment, such as that described by Deng at al\cite{deng2018triplet}, allow the incorporation of ranking information into the training procedure, which may be important in tasks such as gene expression where perturbation signals may be small relative to baseline state. In addition, the field of rank-based embedding learning is intertwined with a broader literature of uni-modal embedding learning, which pioneered such architectures as Twin \cite{koch2015siamese, filzen2017representing} and Triplet networks \cite{hoffer2015deep}, which optimize embeddings to bring similar data together while driving dissimilar data apart. An analysis of best practices of these architecture can be found in Wu et al\cite{wu2017sampling}.



\section{Methods}
\subsection{Dataset \& Tasks}

\noindent \textbf{Data Acquisition and Subsetting}
All data in this study comes from the LINCS Consortium/NIH Next-Generation Connectivity Map Level 3 L1000 data.\cite{subramanian2017next}
This dataset features 978-dimensional gene expression profiles from a variety of human cell lines treated with chemical and genetic perturbations. To ensure support over possible drugs, our data cut uses the most frequent 8 cell lines split into train, validation, and test sets such that no cell line or drug in the train set appears in the validation or test sets. To mitigate non-random missingness, we included only drugs assayed in all cell lines, and limited experiments to those incubated with small molecules for 24 hours at a dose of 10$\mu m$. Final statistics of these data are shown in Table~\ref{tab:dataset_stats}. For drug structures, we used the SMILES \cite{weininger1989smiles} structures provided by LINCS, canonicalized using RDKIT \cite{landrum2006rdkit}.


\noindent \textbf{Preprocessing and Feature Engineering}
Gene expression intensity values from the training, validation, and test sets were centered and scaled at the gene-level based on the mean and standard deviation of each gene intensity across the training set. We  augmented each gene expression profile with three additional sets of features: the corresponding gene expression intensities from a control signature on the same plate, the log$_2$ fold-change between the perturbation and control signatures, and the difference between these gene expression signatures.
%
For use in our baseline and oracle models, we also computed numerical representations of each small molecule: Morgan extended-connectivity fingerprints\cite{rogers2010extended} and the output of the ChemProp network.\cite{yang2019analyzing}

\noindent \textbf{Detailed Task Description}
Our goal is to learn embedders which map molecular structures and gene expression profiles into a vector space such molecular structure embeddings are close to the gene expression profile they induce while being far from other gene expression profiles (Figure 1). Formally, given a collection of gene expression signatures $\mathcal{G}$, chemical structures $\mathcal{M}$, and similarity function $\mathrm{Sim}:\R^d \times \R^d \rightarrow \R$, we seek to learn a gene expression embedder $E_g:\mathcal{G} \rightarrow \R^d$ and chemical embedder $E_m:\mathcal{M} \rightarrow \R^d$ to maximize $\mathrm{Sim}(E_g(g_i), E_m(m_j))$ while simultaneously minimizing $\mathrm{Sim}(E_g(g_i), E_m(m_{\neg j}))$, where gene expression $g_i$ was induced by molecule $m_j$. Unless otherwise specified, the similarity function can be assumed to be Pearson Correlation in our experiments. Across our baseline and oracle methods, we realize many variants of $E_g$ and $E_m$.

\subsection{Baseline and Oracle Methods}
\label{section:baselineMethod}
\noindent \textbf{Nearest Neighbor Baseline}
Nearest-neighbor (NN) methods have been previously shown to establish strong baselines for machine learning tasks on the L1000 data \cite{hodos2017cell, mcdermott2019deep}. In our cross-modal, information retrieval (IR) context, traditional NN methods are not applicable, so we employ the following ``double NN'' baseline: given a gene expression profile as a query, we first identify the nearest gene expression profile in the train set and look up its corresponding small molecule. We then take this small molecule (from the train set) as a query, and return the most structurally similar drug from the \textit{test} set as our final prediction.

In particular, given a mapping $G2M : \mathcal{G} \rightarrow \mathcal{M}$ from gene expression profiles to the small molecule that induced them, and a molecular embedding $E_m$ (which may include molecular fingerprints, Chemprop embeddings, or embeddings learned from other models), we define embedder $E_g: g_\text{query} \mapsto E_m(G2M(\argmax_{g_\text{tr} \in \mathcal G_\text{train}} \mathrm{Sim}(g_\text{query}, g_\text{tr})))$. Then, we perform information retrieval (IR) analyses with such embedders as usual.



\noindent\textbf{Canonical Correlation Analysis Baseline}
Given training matrices of transcriptional $\mathcal{G}_{\text{train}}$ and molecular $\mathcal{M}_{\text{train}}$ encodings, we can learn a set of linear mappings $E_g : \mathcal G_\text{train} \to \mathbb{R}^d$ and $E_m: \mathcal M_\text{train} \to \mathbb{R}^d$ via $d$-dimensional CCA such that these mappings optimize the correlation between elements of $\mathcal G_\text{train}$ and $\mathcal M_\text{train}$.

Note that this procedure requires a default numerical representation for molecules, which, as with other methods, can be either fingerprints, ChemProp embeddings, or learned embeddings by our learning model (described below). CCA can also be performed atop other embedding systems to further optimize embedding results. CCA was performed using SciKit Learn \cite{pedregosa2011scikit}, using 50 components, chosen to optimize validation set performance via a grid-search over a range of 5-125 components, run for 1000 iterations to ensure convergence.


\noindent \textbf{Oracle Models}
The central objective of our task is to learn small molecule embeddings that can stand in as surrogates for their corresponding gene expression signatures. To provide a rough upper-bound for expected performance on this task, we also implemented two ``oracle" models, each of which queries test set GE signatures against pseudo-``chemical embeddings" that are in reality the average GE signatures from each test set drug when it was measured on either (1) the \emph{train set cell lines},\footnote{Note that we can do this as we limited our choice of drugs to those that \emph{were} measured in all 8 cell lines, even though our actual data split prohibits training on any drug that appears in the validation or test sets.} to simulate an embedder that perfectly associates all structures to perturbational profiles, but cannot generalize beyond the train set cell lines, or (2) the \emph{test set cell lines}, which simulates a model of the same capabilities but able to generalize perfectly to the test set as well. These oracle models are still dependent on the underlying gene expression signature representation, so further innovation could offer improved upper bounds for this task. Formally, given $G2M$ mapping gene expression profiles to their corresponding perturbing molecule, we define oracle embeddings $E_g^\text{train}: g_\text{query} \mapsto \mathrm{Avg}(\{g_i \in \mathcal G_\text{train} | G2M(g_i) = G2M(g_\text{query})\})$, and $E_g^\text{test}: g_\text{query} \mapsto \mathrm{Avg}(\{g_i \in \mathcal G_\text{test} | G2M(g_i) = G2M(g_\text{query})\})$. 


\subsection{Deep Coordinated Metric Learning Approach}
\label{section:alignedMethodDescription}


For our learned model, we realize $E_g$ as a self-normalizing neural network (of size dictated by hyperparameter search), and $E_m$ as a directed message-passing neural network (D-MPNN), initialized by the Chemprop system, followed by a feed-forward output layer whose shape was dictated via hyperparameter search.\cite{yang2019analyzing} To train these architectures, we use a margin-based quadruplet loss, building on Wu et al's adaptive margin loss.\cite{wu2017sampling} The base of the adapted margin loss is defined over two data points $i$ and $j$ as $\mathrm{mar}_{\alpha, \beta} := \left( \alpha+y_{i,j} (D_{i j}-\beta) \right)_{+}$,
where $D$ is distance function (here euclidean distance), $\alpha$ defines a permissible margin of separation, $\beta$ controls the boundary between positive and negative pairs, and $y_{i,j}$ is an indicator variable equal to 1 if $i$ and $j$ are of the same class and 0 otherwise ($\alpha$ and $\beta$ were tuned as hyperparameters).

Given two pairs of matching gene expression and molecular structure embeddings, $(g_A, m_A), (g_B, m_B)$, our quadruplet loss is defined as the sum of the margin losses between all cross-modality pairs of embeddings: $\ell_\text{quad} = \mathrm{mar}_{\alpha, \beta}(g_A, m_A) + \mathrm{mar}_{\alpha, \beta}(g_A, m_B) + \mathrm{mar}_{\alpha, \beta}(g_B, m_A) + \mathrm{mar}_{\alpha, \beta}(g_B, m_B)$.
%
The network is thus optimized to bring the positive embedding within the margin of the anchor and negative embedding outside the margin.
%
%
For sampling these two pairs (an analog of negative sampling for a more traditional triplet network), we first sample one matching pair, choose the molecular structure for the other pair based on the distance-weighted negative sampling scheme described in Wu et al, which was successful with their margin-based approach \cite{wu2017sampling}, then fill in the other gene expression profile to match the sampled molecular structure. To make this process computationally efficient we pre-computed the average distance in \emph{average post-perturbational gene expression space} between every pair of small molecule structures in the dataset. 
We additionally tried other losses, including two varieties of traditional triplet losses, and a quintuplet loss, but ultimately found the quadruplet loss to be most performant via our hyperparameter search.
 
\noindent \textbf{Training and Hyperparameter Selection}
\label{sec:hyptun}
Each model was trained on a Nvidia GeForce GTX 1080 GPU. Early stopping was used to select the model with the best mean reciprocal rank on the validation set. Hyperparameter tuning via the Bayesian Hyperopt library~\cite{bergstra2013making} was performed over a wide range of possible hyperparameters, including network depth and width (parametrized by the size of the first hidden layer and a growth rate), learning rate, number of epochs, batch size, margin and $\beta$ parameters, triplet model orientation (i.e., gene first or compound first), activation function/network type (e.g., SNNN vs. unconstrained fully connected network), and dropout, with no early stopping for hyperparameter search runs. The optimal hyperparameters from this search are shown in Appendix Section~\ref{sec:final_params}.

\subsection{Experiments}

We designed a range of experiments with two purposes: First, we sought to evaluate if and how our deep coordinated representation learning method offers improvements over principled baselines. This entails a quantitative performance comparison against baseline methods and ablated versions of our model. Second, we introspect into the representations learned by training on this new task, to better understand the challenges and utility of the general framework. This entails a quantitative performance comparison against oracle models, a statistical analysis probing the ability of our models to generalize to new structures, and a qualitative exploration of the changes in chemical representations that are induced by our training scheme.

Quantitative performance analyses began by computing the embeddings of gene expression signatures and chemical structures in the test set, using the baseline, oracle, and deep coordinated methods defined in Sections~\ref{section:baselineMethod} and ~\ref{section:alignedMethodDescription}. Using each embedded gene expression signature as a query, we ranked all chemical structures in the test set based on their proximity to that gene expression signature in the embedding space. These rankings were then used to compute standard information-retrieval metrics (precision-recall curves, MR, MRR, and Hits at/H@ 10 or 100). For our ablation analyses, we repeated the above experiments using various combinations of raw and learned GE and chemical representations.

In addition to the information retrieval analyses, we probed the generalizability of our representations by analysing the statistical relationship between the average retrieval performance for each chemical structure and the structural similarity to the most similar chemical in the training set. Similarity was measured via the Tanimoto distance between all pairs of molecular fingerprints in the train and test set. We further examined performance vs. chemical specificity, using the number of genes that a molecule, on average, affected as a measure of specificity. Finally, we visualized the latent space of our chemical embedder (versus their pre-trained representations learned by ChemProp), and noted the relative position of drugs with the same mechanism of action (MOA) in each latent space.

\section{Results \& Discussion}

\subsection{Quantitative IR Experiments}

\begin{figure}
\centering
\includegraphics[width=0.55\linewidth]{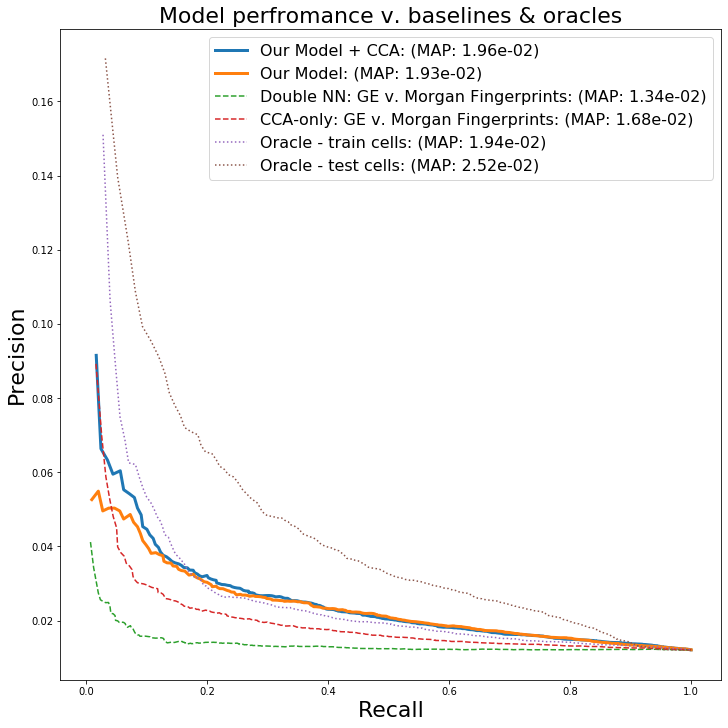}
\caption{Precision Recall curves for drug identification given gene expression signatures, across various baselines (dashed lines), oracles (dotted lines) and our model (solid lines).}
\label{fig:PRCurves}
\end{figure}

\textbf{Baseline and Proposed Method} Information retrieval results from the baseline and proposed methods are reported in Figure~\ref{fig:PRCurves}. This figure shows that our model variants offer significant performance improvements over either of the CCA or NN baselines, and even approach the performance of the train-set only oracle model. All models fall dramatically short of the test-set generalizability oracle, which indicates that while our tasks offer significant improvement over baseline models here, there are still major gains possible, primarily by focusing on improving the generalizability to the novel cell-types of the test set.

In addition, we show various ablation studies over the baseline models in Table~\ref{tab:main_results} to probe what gene or molecular representations would make them better or worse. Strikingly, we can note that uniformly using aligned representations (meaning representations based on our multi-view alignment neural network architecture) offers significant improvements over other representations, indicating that even with a baseline approach such as the double nearest neighbor (D-NN) model, improvements to the embedding quality translate to notable improvements to IR performance. Notably, this is true both for GE and Chemical embedders, with Aligned-Aligned representations yielding optimal performance for both D-NN and CCA query mechanisms. Additionally, it is also clear that CCA is the preferred query metric, over either raw correlation (Corr) based lookups or D-NN based lookups.

\begin{table}
\tbl{
IR metrics across various configurations of the model/baselines. `Chemprop' refers to pretrained model from Yang et al \cite{yang2019analyzing}. `Aligned' indicates representations learned from our method (see Section~\ref{section:alignedMethodDescription}). MR=median rank, MRR=mean reciprocal rank, H@K=Hit/Recall at K.
}{
\begin{tabular}{lllrrrr}
\toprule
\textbf{GE}          &  \textbf{Chemical}   & \textbf{Method} &  \textbf{MR}  & \textbf{MRR}   & \textbf{H@10}  & \textbf{H@100} \\
\midrule 
Raw         & Morgan FP  & D-NN  & 206 & 0.025   & 0.037 & 0.240 \\
Raw         & Chemprop   & D-NN   & 211 & 0.025   & 0.035 & 0.254 \\
Raw         & Aligned & D-NN    & 189 & 0.033   & 0.047 & 0.290 \\
Aligned       & Morgan FP  & D-NN   & 214 & 0.025   & 0.041 & 0.256 \\
Aligned   & Chemprop   & D-NN    & 196 & 0.022  &  0.037 & 0.278 \\
Aligned         & Aligned & D-NN    & 137 & 0.039   & 0.072 & 0.402 \\
\midrule
Raw         & Morgan FP  & CCA   & 180 & 0.027  & 0.045 & 0.303 \\
Raw         & Chemprop   & CCA   & 184 & 0.024  & 0.040 & 0.294 \\ 
Raw         & Aligned & CCA   & 134 & 0.039   & 0.076 & 0.412 \\
Aligned  & Morgan FP  & CCA   & 177 & 0.027  & 0.050 & 0.319 \\ 
Aligned  & Chemprop   & CCA   & 163 & 0.028  & 0.051 & 0.334 \\
Aligned  & Aligned & CCA   & 130 & \textbf{0.048}  &  \textbf{0.093} & 0.425 \\
\midrule
Aligned  & Aligned & Corr  & \textbf{126} & 0.042  & 0.085 &\textbf{ 0.432} \\ 
\bottomrule
\end{tabular}}
\label{tab:main_results}
\end{table}

\textbf{Oracle Model Analysis} Results from the oracle models are reported in Figure~\ref{fig:eCDF} and Table~\ref{tab:oracle_results}. As expected, oracle models using GE signatures from the test cell line greatly outperformed those using signatures from the train cell lines. This stark difference suggests that one of the largest barriers to performance here is the generalization gap between different cell lines. This further motivates for the curation of larger, cell-line heterogeneous datasets in the future.

In addition, aligned embeddings modestly improved the performance of test set oracles, and greatly improved the performance of train set oracles, consistent with the GE embedders learning slightly more generalizable representations. Of note, our proposed model achieved comparable results as the oracle model that leveraged raw GE signatures from the train cell lines. More specifically, our approach yielded slightly worse than the oracle on metrics (MRR, H@10) that emphasize early rankings, and slightly better on metrics (MR, H@100) that focus on more aggregate results. This is also apparent in the precision-recall plot, which shows the aligned embeddings curve starting out slightly below that of the raw/train oracle curve but then moving rapidly above it as further results are considered.

\begin{table}
\tbl{
IR metrics for the various oracle methods.
}{
\begin{tabular}{lrrrr}
\toprule
\textbf{Oracle Model} & \textbf{MR}  & \textbf{MRR}   & \textbf{H@10}  & \textbf{H@100} \\ 
\midrule
Oracle - Train Cell Lines (Raw GE)   & 138 & 0.057 & 0.101 & 0.400 \\ 
Oracle - Train Cell Lines (Embed)   & 110 & 0.064 & 0.128 & 0.466 \\ 
Oracle - Test Cell Line (Raw GE)   & 82 & 0.076 & 0.147 & 0.565 \\ 
Oracle - Test Cell Line (Embed)   & 79 & 0.093 & 0.160 & 0.568 \\ 
\midrule
Our Approach   &  126 & 0.042 & 0.085 & 0.432 \\ 

\bottomrule
\end{tabular}}
\label{tab:oracle_results}
\end{table}

\subsection{Introspection Analyses}
\begin{figure}
    \centering
    \includegraphics[width=0.9\linewidth]{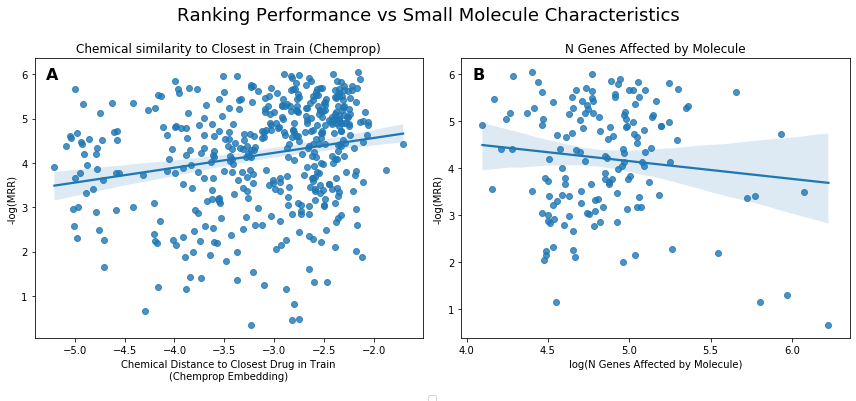}
    \caption{Left: Performance (lower is better) vs. structural distance to the nearest compound in the training set. This plot demonstrates that compounds more structurally dissimilar to the train set show mildly worse performance than those that are more similar. See Appendix Figure~\ref{fig:performance_predictors_append} for analogous plots for four additional measures of distance from the training set. Right: Average Performance vs. \# of Genes deferentially expression following treatment with the molecule of interest, showing that compounds that have broader transcriptomic effects are better retrieved by this method.}
    \label{fig:perf_predictors}
\end{figure}

%
Figure~\ref{fig:perf_predictors}A 
contains the results of our experiments comparing performance to distance from the training set. Regardless of the measure of chemical similarity, compound retrieval performance was inversely correlated with distance from the training set. As can be seen in Supplementary Figure~\ref{fig:performance_predictors_append}, the same trend held with learned gene expression embeddings, and was present but much weaker using raw gene expression profiles.

Figure~\ref{fig:perf_predictors}B depicts the relationship between the transcriptional specificity of a compound and its ability to be retrieved using our analysis. As can be seen, there is a mild negative correlation, implying that molecules that affect the expression of many genes are easier to retrieve using this approach. Note that this observation is concordant with our findings on the difficulty of generalizing to new cell lines -- drugs that affect a small, targeted set of genes are more likely to be cell line specific, and as our model is forced to surmount a significant generalization gap in evaluation, such cell-line specific signals are largely wiped out.

In addition, our analysis of the changes induced in the embedding space, shown in Supplementary Figure~\ref{fig:clustering_moas}, reveal that our model's embeddings of molecules appear to better cluster shared MOAs than do the raw ChemProp embeddings, from which our model is initialized. This suggests that, as hypothesized regarding the nature of this task, our model is learning rich representations of the underlying molecules, though additional work remains to investigate this effect more thoroughly.






\subsection{Future Work}

We see several opportunities for further work on this task. First, expanding our data coverage, across molecules, cell lines, dosages, and treatment durations will allow us to measure and improve generalizability here. Second, exploring additional strategies to use the over-sampled nature of these data (e.g., ensembling together control and perturbational signatures to reduce variance) could be beneficial. Third, a more robust exploration of model architectures, losses, and deep metric learning/negative sampling methods, could offer improvements here.\cite{li2018survey} Additionally, other styles of multimodal embedding could be explored, such as the use of cycle generative adversarial network, which in particular would enable us to adopt a semi-supervised approach.\cite{felix2018multi,mcdermott_semi-supervised_2018}
The use of interpretability methods, particularly those used for graph analyses,\cite{ying2019gnnexplainer} as well as additional studies interrogating how our model's performance changes with the amount of available training data could also be insightful here. Fourth, we recommend exploring methods to improve cell-line generalizability, e.g. incorporating information across many cell lines when forming predictions. Finally, we also note that while our analyses only examine small-molecule therapeutics, similar methods could also be applied to other modalities, such as RNA-based therapies.

\section{Conclusion}

We present a new task: cross-modal multi-view alignment between drug structures and perturbational gene expression profiles, which links molecular structure to an objective, functional readout of drugs with very broad biomedical relevance. We profile state-of-the-art representation learning methods on this new task, and inspect the learned chemical embeddings. We find that this modeling task induces an embedding space reflective of drug mechanism of action -- which is not explicitly included in the training regime -- and see modest generalization to both new structures and a new biological environment. Our oracle experiments demonstrate major performance gaps when trying to generalize to new tissues. We hope that this new benchmark task will catalyze future research and ultimately help enable a rapid, in silico compound prioritization methods.

\section{Acknowledgements}
The authors thank Connor Coley and Kyle Swanson for providing the pretrained chemical embedder from Yang et al \cite{yang2019analyzing}. S.G.F. was supported by training grant T32GM007753 from the National Institute of General Medical Science. M.B.A.M. was funded in part by National Institutes of Health: National Institutes of Mental Health grant P50-MH106933 as well as a Mitacs Globalink Research Award. The content is solely the responsibility of the authors.

\bibliographystyle{ws-procs11x85}
\bibliography{small_mol_align}

\begin{thebibliography}{10}

\bibitem{hansson2002molecular}
T.~Hansson, C.~Oostenbrink and W.~van Gunsteren, Molecular dynamics
  simulations, {\em Current opinion in structural biology} {\bf 12}, 190
  (2002).

\bibitem{krallinger2017information}
M.~Krallinger, O.~Rabal, A.~Lourenco, J.~Oyarzabal and A.~Valencia, Information
  retrieval and text mining technologies for chemistry, {\em Chemical reviews}
  {\bf 117}, 7673  (2017).

\bibitem{battaglia2018relational}
P.~W. Battaglia, J.~B. Hamrick, V.~Bapst, A.~Sanchez-Gonzalez, V.~Zambaldi,
  M.~Malinowski, A.~Tacchetti, D.~Raposo, A.~Santoro, R.~Faulkner {\em et~al.},
  Relational inductive biases, deep learning, and graph networks, {\em arXiv
  preprint arXiv:1806.01261}   (2018).

\bibitem{wu2018moleculenet}
Z.~Wu, B.~Ramsundar, E.~N. Feinberg, J.~Gomes, C.~Geniesse, A.~S. Pappu,
  K.~Leswing and V.~Pande, Moleculenet: a benchmark for molecular machine
  learning, {\em Chemical science} {\bf 9}, 513  (2018).

\bibitem{kearnes2016molecular}
S.~Kearnes, K.~McCloskey, M.~Berndl, V.~Pande and P.~Riley, Molecular graph
  convolutions: moving beyond fingerprints, {\em Journal of computer-aided
  molecular design} {\bf 30}, 595  (2016).

\bibitem{yang2019analyzing}
K.~Yang, K.~Swanson, W.~Jin, C.~W. Coley, P.~Eiden, H.~Gao, A.~Guzman-Perez,
  T.~Hopper, B.~Kelley, M.~Mathea {\em et~al.}, Analyzing learned molecular
  representations for property prediction, {\em Journal of chemical information
  and modeling}   (2019).

\bibitem{lo2018machine}
Y.-C. Lo, S.~E. Rensi, W.~Torng and R.~B. Altman, Machine learning in
  chemoinformatics and drug discovery, {\em Drug discovery today} {\bf 23},
  1538  (2018).

\bibitem{ramsundar2015massively}
B.~Ramsundar, S.~Kearnes, P.~Riley, D.~Webster, D.~Konerding and V.~Pande,
  Massively multitask networks for drug discovery, {\em arXiv preprint
  arXiv:1502.02072}   (2015).

\bibitem{faber2017prediction}
F.~A. Faber, L.~Hutchison, B.~Huang, J.~Gilmer, S.~S. Schoenholz, G.~E. Dahl,
  O.~Vinyals, S.~Kearnes, P.~F. Riley and O.~A. Von~Lilienfeld, Prediction
  errors of molecular machine learning models lower than hybrid dft error, {\em
  Journal of chemical theory and computation} {\bf 13}, 5255  (2017).

\bibitem{gilmer2017neural}
J.~Gilmer, S.~S. Schoenholz, P.~F. Riley, O.~Vinyals and G.~E. Dahl, Neural
  message passing for quantum chemistry, {\em Proceedings of the 34th
  International Conference on Machine Learning-Volume 70} , 1263  (2017).

\bibitem{zitnik2018modeling}
M.~Zitnik, M.~Agrawal and J.~Leskovec, Modeling polypharmacy side effects with
  graph convolutional networks, {\em Bioinformatics} {\bf 34}, i457  (2018).

\bibitem{musa2017review}
A.~Musa, L.~S. Ghoraie, S.-D. Zhang, G.~Glazko, O.~Yli-Harja, M.~Dehmer,
  B.~Haibe-Kains and F.~Emmert-Streib, A review of connectivity map and
  computational approaches in pharmacogenomics, {\em Briefings in
  bioinformatics} {\bf 19}, 506  (2017).

\bibitem{lamb2006connectivity}
J.~Lamb, E.~D. Crawford, D.~Peck, J.~W. Modell, I.~C. Blat, M.~J. Wrobel,
  J.~Lerner, J.-P. Brunet, A.~Subramanian, K.~N. Ross {\em et~al.}, The
  connectivity map: using gene-expression signatures to connect small
  molecules, genes, and disease, {\em science} {\bf 313}, 1929  (2006).

\bibitem{subramanian2017next}
A.~Subramanian, R.~Narayan, S.~M. Corsello, D.~D. Peck, T.~E. Natoli, X.~Lu,
  J.~Gould, J.~F. Davis, A.~A. Tubelli, J.~K. Asiedu {\em et~al.}, A next
  generation connectivity map: L1000 platform and the first 1,000,000 profiles,
  {\em Cell} {\bf 171}, 1437  (2017).

\bibitem{musa2018harnessing}
A.~Musa, S.~Tripathi, M.~Kandhavelu, M.~Dehmer and F.~Emmert-Streib, Harnessing
  the biological complexity of big data from lincs gene expression signatures,
  {\em PloS one} {\bf 13}, p. e0201937  (2018).

\bibitem{liu2018systematic}
T.-P. Liu, Y.-Y. Hsieh, C.-J. Chou and P.-M. Yang, Systematic polypharmacology
  and drug repurposing via an integrated l1000-based connectivity map database
  mining, {\em Royal Society open science} {\bf 5}, p. 181321  (2018).

\bibitem{clark2014characteristic}
N.~R. Clark, K.~S. Hu, A.~S. Feldmann, Y.~Kou, E.~Y. Chen, Q.~Duan and
  A.~Ma’ayan, The characteristic direction: a geometrical approach to
  identify differentially expressed genes, {\em BMC bioinformatics} {\bf 15},
  p.~79  (2014).

\bibitem{donner2018drug}
Y.~Donner, S.~Kazmierczak and K.~Fortney, Drug repurposing using deep
  embeddings of gene expression profiles, {\em Molecular pharmaceutics} {\bf
  15}, 4314  (2018).

\bibitem{dincer2018deepprofile}
A.~B. Dincer, S.~Celik, N.~Hiranuma and S.-I. Lee, Deepprofile: Deep learning
  of cancer molecular profiles for precision medicine, {\em bioRxiv} , p.
  278739  (2018).

\bibitem{rampavsek2019dr}
L.~Ramp{\'a}{\v{s}}ek, D.~Hidru, P.~Smirnov, B.~Haibe-Kains and A.~Goldenberg,
  Dr. vae: improving drug response prediction via modeling of drug perturbation
  effects, {\em Bioinformatics} {\bf 35}, 3743  (2019).

\bibitem{cheng2013evaluation}
J.~Cheng, Q.~Xie, V.~Kumar, M.~Hurle, J.~M. Freudenberg, L.~Yang and
  P.~Agarwal, Evaluation of analytical methods for connectivity map data (World
  Scientific, 2013) pp. 5--16.

\bibitem{lopez2018deep}
R.~Lopez, J.~Regier, M.~B. Cole, M.~I. Jordan and N.~Yosef, Deep generative
  modeling for single-cell transcriptomics, {\em Nature methods} {\bf 15}, 1053
   (2018).

\bibitem{hsu2018unsupervised}
T.-M.~H. Hsu, W.-H. Weng, W.~Boag, M.~McDermott and P.~Szolovits, Unsupervised
  multimodal representation learning across medical images and reports, {\em
  arXiv preprint arXiv:1811.08615}   (2018).

\bibitem{hassani2020contrastive}
K.~Hassani and A.~H. Khasahmadi, Contrastive multi-view representation learning
  on graphs, {\em arXiv preprint arXiv:2006.05582}   (2020).

\bibitem{hotelling1992relations}
H.~Hotelling, Relations between two sets of variates (Springer, 1992) pp.
  162--190.

\bibitem{li2018survey}
Y.~Li, M.~Yang and Z.~M. Zhang, A survey of multi-view representation learning,
  {\em IEEE Transactions on Knowledge and Data Engineering}   (2018).

\bibitem{deng2018triplet}
C.~Deng, Z.~Chen, X.~Liu, X.~Gao and D.~Tao, Triplet-based deep hashing network
  for cross-modal retrieval, {\em IEEE Transactions on Image Processing} {\bf
  27}, 3893  (2018).

\bibitem{koch2015siamese}
G.~Koch, R.~Zemel and R.~Salakhutdinov, Siamese neural networks for one-shot
  image recognition, {\em ICML deep learning workshop} {\bf 2}  (2015).

\bibitem{filzen2017representing}
T.~M. Filzen, P.~S. Kutchukian, J.~D. Hermes, J.~Li and M.~Tudor, Representing
  high throughput expression profiles via perturbation barcodes reveals
  compound targets, {\em PLoS computational biology} {\bf 13}, p. e1005335
  (2017).

\bibitem{hoffer2015deep}
E.~Hoffer and N.~Ailon, Deep metric learning using triplet network, {\em
  International Workshop on Similarity-Based Pattern Recognition} , 84  (2015).

\bibitem{wu2017sampling}
C.-Y. Wu, R.~Manmatha, A.~J. Smola and P.~Krahenbuhl, Sampling matters in deep
  embedding learning, {\em Proceedings of the IEEE International Conference on
  Computer Vision} , 2840  (2017).

\bibitem{weininger1989smiles}
D.~Weininger, A.~Weininger and J.~L. Weininger, Smiles. 2. algorithm for
  generation of unique smiles notation, {\em Journal of chemical information
  and computer sciences} {\bf 29}, 97  (1989).

\bibitem{landrum2006rdkit}
G.~Landrum {\em et~al.}, Rdkit: Open-source cheminformatics  (2006).

\bibitem{rogers2010extended}
D.~Rogers and M.~Hahn, Extended-connectivity fingerprints, {\em Journal of
  chemical information and modeling} {\bf 50}, 742  (2010).

\bibitem{hodos2017cell}
R.~Hodos, P.~Zhang, H.-C. Lee, Q.~Duan, Z.~Wang, N.~R. Clark, A.~Ma’ayan,
  F.~Wang, B.~Kidd, J.~Hu {\em et~al.}, Cell-specific prediction and
  application of drug-induced gene expression profiles, {\em Pacific Symposium
  on Biocomputing} {\bf 23}  (2017).

\bibitem{mcdermott2019deep}
M.~McDermott, J.~Wang, W.~N. Zhao, S.~D. Sheridan, P.~Szolovits, I.~Kohane,
  S.~J. Haggarty and R.~H. Perlis, Deep learning benchmarks on l1000 gene
  expression data, {\em IEEE/ACM transactions on computational biology and
  bioinformatics}   (2019).

\bibitem{pedregosa2011scikit}
F.~Pedregosa, G.~Varoquaux, A.~Gramfort, V.~Michel, B.~Thirion, O.~Grisel,
  M.~Blondel, P.~Prettenhofer, R.~Weiss, V.~Dubourg {\em et~al.}, Scikit-learn:
  Machine learning in python, {\em Journal of machine learning research} {\bf
  12}, 2825  (2011).

\bibitem{bergstra2013making}
J.~Bergstra, D.~Yamins and D.~D. Cox, Making a science of model search:
  Hyperparameter optimization in hundreds of dimensions for vision
  architectures, {\em Proceedings of the 30th International Conference on
  International Conference on Machine Learning - Volume 28} , p.
  I–115–I–123  (2013).

\bibitem{felix2018multi}
R.~Felix, V.~B. Kumar, I.~Reid and G.~Carneiro, Multi-modal cycle-consistent
  generalized zero-shot learning, {\em Proceedings of the European Conference
  on Computer Vision (ECCV)} , 21  (2018).

\bibitem{mcdermott_semi-supervised_2018}
M.~B.~A. McDermott, T.~Yan, T.~Naumann, N.~Hunt, H.~Suresh, P.~Szolovits and
  M.~Ghassemi, Semi-{Supervised} {Biomedical} {Translation} with {Cycle}
  {Wasserstein} {Regression} {GANs}, p.~8  (2018).

\bibitem{ying2019gnnexplainer}
Z.~Ying, D.~Bourgeois, J.~You, M.~Zitnik and J.~Leskovec, Gnnexplainer:
  Generating explanations for graph neural networks, {\em Advances in neural
  information processing systems} , 9244  (2019).

\end{thebibliography}

\clearpage
\appendix

\section{Appendix}

\subsection{Data Access}

Raw LINCS data are available for download at \url{https://www.ncbi.nlm.nih.gov/geo/query/acc.cgi?acc=GSE70138} and \url{https://www.ncbi.nlm.nih.gov/geo/query/acc.cgi?acc=GSE92742}. Our preprocessed versions of these files will be distributed with our Github repository.

\subsection{Final Optimal Parameters}
\label{sec:final_params}

\begin{table}
\centering
\tbl{
    Dataset cell lines, number of samples, and number of drugs per split.
}{
    \begin{tabular}{lp{0.4\linewidth}ll} \toprule
                            & Train & Validation & Test \\ \midrule
        Cell Lines          & HEPG2, HA1E, HCC515, VCAP, A375, PC3, MCF7& A549       & HT29 \\
        \# Samples          & 97210 & 1255       & 2317\\
        \# Drugs            & 3490  & 436        & 437 \\
        \bottomrule
    \end{tabular}
}
\label{tab:dataset_stats}
\end{table}

Final, optimal hyperparameters were selected according to the median rank of the parameter setting. Final values are shown in Table~\ref{tab:hyperparams}. We see lots of consistency between the two settings, despite the fact that both searches were independent. For example, both prefer very similar gene expression architectures, both run for 12 epochs, and both have relatively low margins.

\begin{table}[h!]
    \centering
    \tbl{Final optimal hyperparameters for the deep neural network architecture.}{
    \begin{tabular}{lrr} \toprule
        Parameter            & Drug Split \\ \midrule
        \# Epochs            & 12         \\
        LR                   & 5.6e-4     \\
        Batch Size           & 417        \\
        Rank Transform       & No         \\
        Neg. Samp.           & N/A        \\
        Gene Exp. Emb. Arch. & 677, 1159  \\
        Embed Size           & 51       \\
        Dropout Prob         & 0.02       \\
        Activation           & SELU       \\
        Linear Bias          & False      \\
        Beta                 & 1.3        \\
        Margin               & 0.31       \\
        RdKit Ft.            & True       \\
    \bottomrule \end{tabular}}
    \label{tab:hyperparams}
\end{table}



\begin{figure}
\centering
\includegraphics[width=\linewidth]{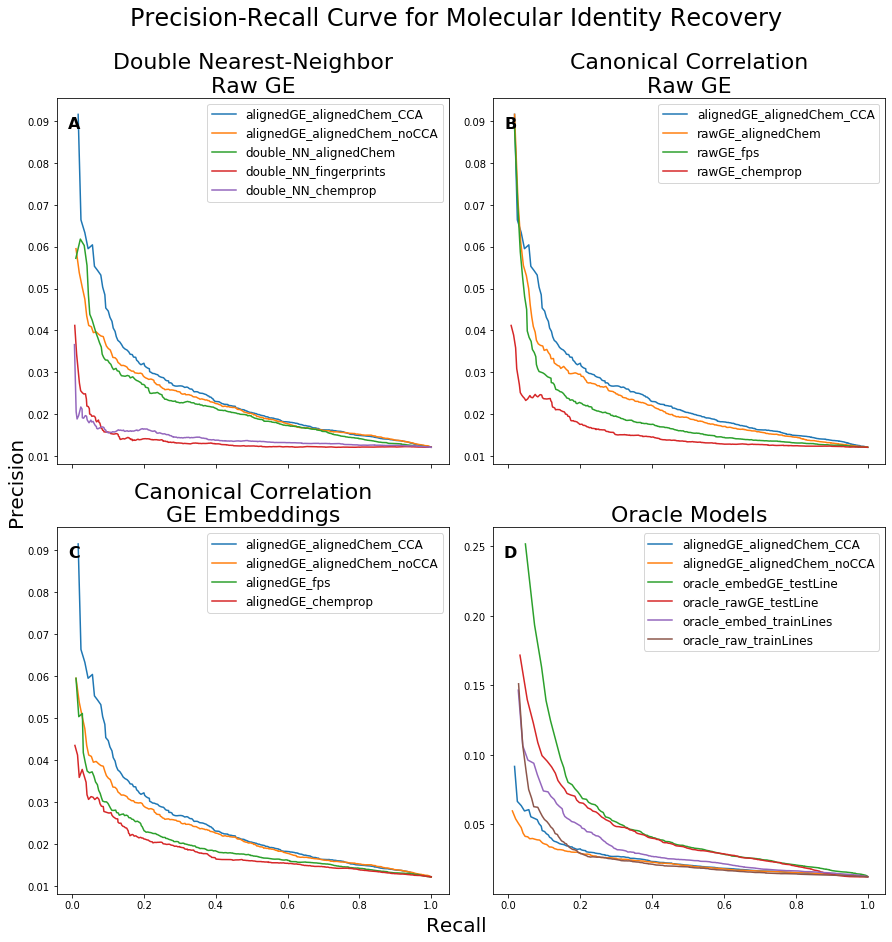}
\caption{Precision Recall curves for drug identification given gene expression signatures. The ``alignedGE\_alignedChem\_CCA" and ``alignedGE\_alignedChem\_noCCA" curves, plotted in blue and orange, respectively, in all plots, represents our approach as described in Section~\ref{section:alignedMethodDescription} with and without CCA post-processing. All other methods represent baseline or oracle methods as described in Section~\ref{section:baselineMethod}.}
\label{fig:eCDF}
\end{figure}


\begin{figure}[!h]
    \centering
    \includegraphics[width=\linewidth]{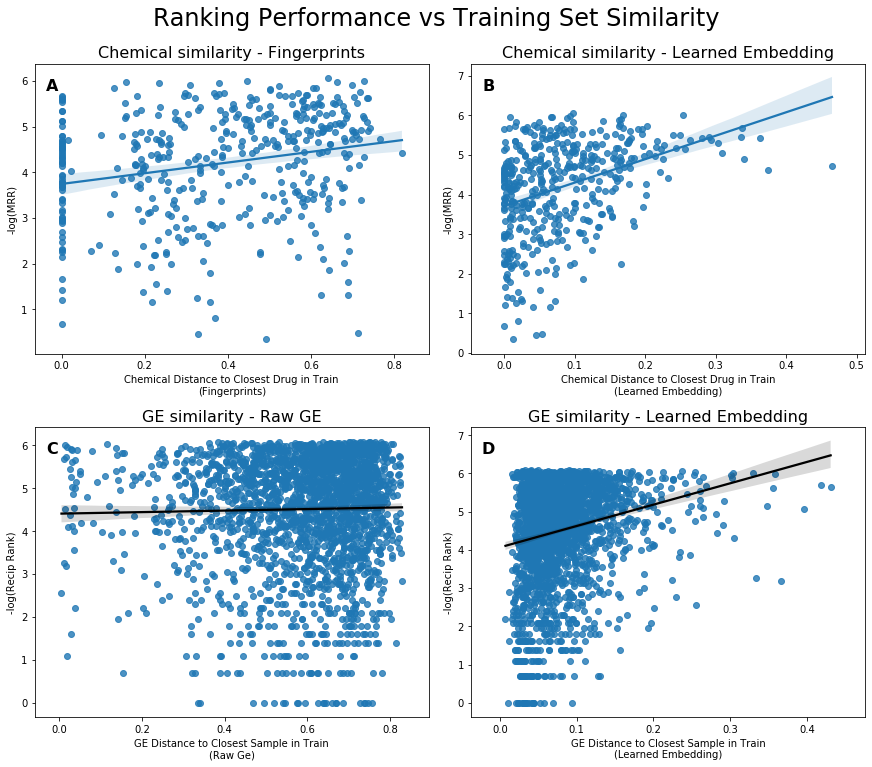}
    \caption{Alternative measures of performance as a function of distance from the training set. Accompanies Figure~\ref{fig:perf_predictors}. Note that while some test-set molecules have tanimoto distance of 0 to the train set, this does not imply they were shared between the two sets; rather, there are cases of two distinct molecules with identical molecular fingerprints. See Figure~\ref{fig:mols_w_small_fingerprints} for several examples.}
    \label{fig:performance_predictors_append}
\end{figure}

\begin{figure}[!h]
    \centering
    \includegraphics[width=0.5\linewidth]{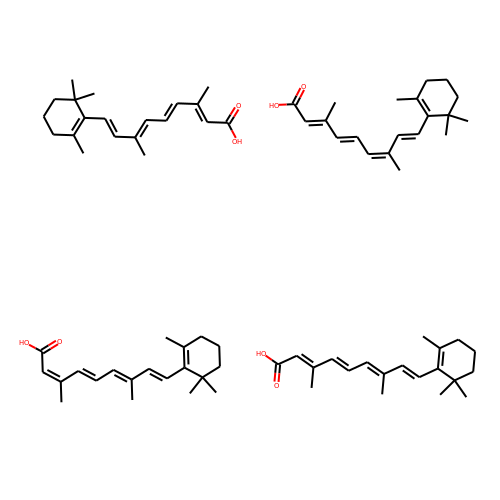}
    \caption{Examples of four molecules that have slightly distinct structures, but have identical Morgan Circular Fingerprints.}
    \label{fig:mols_w_small_fingerprints}
\end{figure}

\begin{figure}
    \centering
    \includegraphics[width=\linewidth]{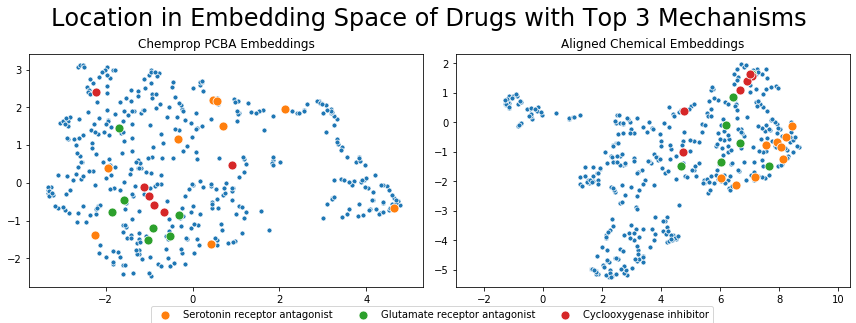}
    \caption[Visualization of Latent Chemical Embedding Space]{This figure shows our induced embeddings (right) compared to established chemical embeddings (left), projected to 2-dimensions using UMAP, colored according to MOA. (Blue dots represent small molecules with different or unknown MOAs.)}
    \label{fig:clustering_moas}
\end{figure}

\begin{figure}
    \centering
    \includegraphics[width=\linewidth]{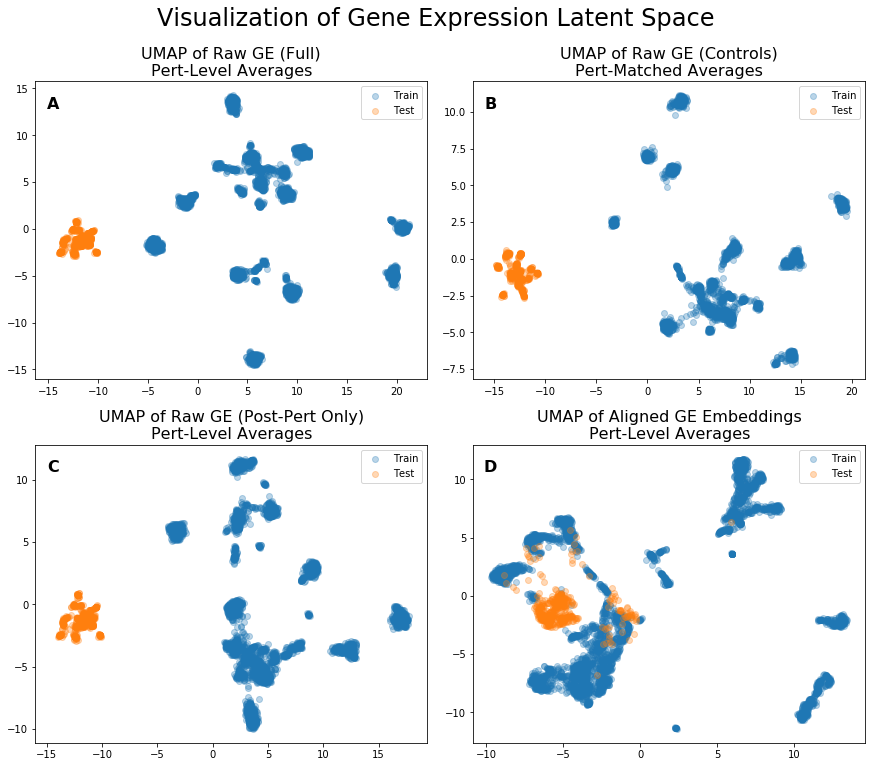}
    \caption{UMAP latent space of gene expression signatures, colored by train vs test set. A: Full raw gene expression signatures (post-perturbational signature, plate control signature, log-fold change). B: Raw gene expression from control signatures. C: Raw gene expression values, only the post-perturbational measurments. D: Aligned embedding representations.}
    \label{fig:clustering_genes_controls}
\end{figure}

\end{document}